# Combining Visual and Photoelectric Observations of Semi-Regular Red Variables

**Terry T. Moon**
*Astronomical Society of South Australia (ASSA) GPO Box 199, Adelaide, SA 5001, Australia; e-mail: terry.moon@bigpond.com*

**Sebastián A. Otero**
*(1) Grupo Wezen 1 88, Buenos Aires, Argentina; e-mail: varsao@fullzero.com.ar*
*(2) Centro de Estudios Astronómicos (CEA), Mar del Plata, Argentina*

**Laszlo L. Kiss**
*School of Physics, University of Sydney, NSW 2006, Australia; e-mail: laszlo@physics.usyd.edu.au*

**Abstract** Combining visual observations of SR variables with measurements of them using a photoelectric photometer is discussed then demonstrated using data obtained for the bright, southern SR variable θ Aps. Combining such observations is useful in that it can provide a more comprehensive set of data by extending the temporal coverage of the light curve. Typically there are systematic differences in the visual and photometric datasets that must be corrected for.

## 1. Introduction

The authors have been undertaking independent observing programs of variable stars (one in Argentina and the other in Australia) that include a group of bright semi-regular red variables (SR) at southern declinations. One of us, Sebastian Otero, is undertaking a program of visual observations using a modified version of the Argelander method (Otero, Fraser and Lloyd 2001) while another, Terry Moon, regularly obtains PEP measurements of about 30 bright SR variables located at southern declinations. A number of these are being monitored by both programs, namely β Gru, θ Aps, X TrA, SX Pav, Y Pav, R Dor, BO Mus, and V744 Cen. The combining of visual estimates of magnitude with PEP measurements is thus of interest as it affords us the opportunity to increase coverage of the light curves for these semi-regular red variables. The importance of the extent of the set of observations was emphasized by Kiss *et al*. (1999).

In an earlier paper, Otero & Moon (2006) combined their independent observations of β Gru to determine its characteristic period of pulsation. For that paper an overlap in the two sets of data was used to evaluate any mean difference between them. Subsequently a small correction of -0.03 magnitude was applied to the visual estimates to bring them into accord with the photoelectric measurements. This paper discusses some of the issues associated with combining visual and photoelectric data, the discrepancies that arise, and why, then illustrates a practical approach to combining such data using observations of θ Aps as an example.

## 2. Visual Estimates versus Photoelectric Measurements

Undertaking visual observations of variable stars remains popular owing to a number of advantages but there are also some significant drawbacks (Simonsen 2004). Henden and Kaitchuck (1990) note that 'The human eye can generally interpolate the brightness of one star relative to nearby comparisons to about 0.2 magnitude.' More recently Toone (2005) has pointed out that there can also be discrepancies of 0.2 magnitudes or more between comparison star sequences from different sources. Using a modified version of the Argelander method where the observer makes allowances for color differences by observing comparison stars encompassing a wide range of spectral types, and uses photoelectric rather than visual sequences of comparison stars, such discrepancies may be reduced to 0.1 magnitudes.

Photoelectric photometry (Henden and Kaitchuck 1990) provides precise measurements of variable stars that can be recorded and then reduced to accurate magnitudes. Using different filters, color indices can also be measured yielding additional information of astrophysical significance. Typically, the precision of PEP measurements is better than 0.01 magnitudes but the scatter in $V$ and $B$-$V$ determined from many observers transforming their measurements to the standard *UBV* system appears to be about 0.02 magnitudes (Böhm-Vitense 1981).

Table 1 summarizes the comparative advantages and disadvantages of visual and PEP observations.

*Table 1. Comparison of visual estimates with PEP measurements.*

|  | **Visual estimates** | **PEP measurements** |
|---|---|---|
| **Advantages** | Minimal equipment and technical training required. | Accurate measurements of magnitudes and color indices to 0.02 mag. or better. |
|  | Results obtained with relative ease not requiring extensive processing of data. | Additional information (color indices) obtained through use of different filters |
|  | Can be undertaken in poorer seeing conditions | Systematic corrections can be applied for color differences and atmospheric extinction. |
| **Disadvantages** | Quality of results is highly dependent on the skill of the observer with precision seldom better than 0.1 magnitudes and the error of an individual observation being typically 0.3 magnitudes. | Requires significant investment in equipment and technical training. |
|  | Significant scope for human bias to be introduced. | Involves significant effort to both obtain the data and process it. |
|  | Limited to visual wavelengths | Highly dependent on seeing conditions |
|  | Difficulty in systematically correcting for color differences between stars and the effects of atmospheric extinction. |  |

## 3. Comparison of Spectral Responses

Figure 1 compares the spectral responses of the dark-adapted human eye (Allen 1973, Cox 2000) and the Optec photometer (Optec 2007) used for the photoelectric measurements presented in this paper, with that of the standard, photoelectric *V*-band (Allen 1973). The spectral response of the Optec *V*-band is sufficiently close to that defined by the standardized *UBV* photometric system so that a linear relation can be applied to transform the measurements made to *V* magnitudes. Measurements of stars

(encompassing a wide range of spectral types) made with this photometer confirm that a linear relation with respect to *B-V* suffices for transforming the PEP *V*-band measurements to standardized *V* magnitudes and, similarly, a linear relationship also suffices for transforming measured to standard *B-V* indices. Figure 1 illustrates that the spectral response of the dark-adapted human eye (scotopic vision), in contrast, varies significantly from that for the standard *V*-band.

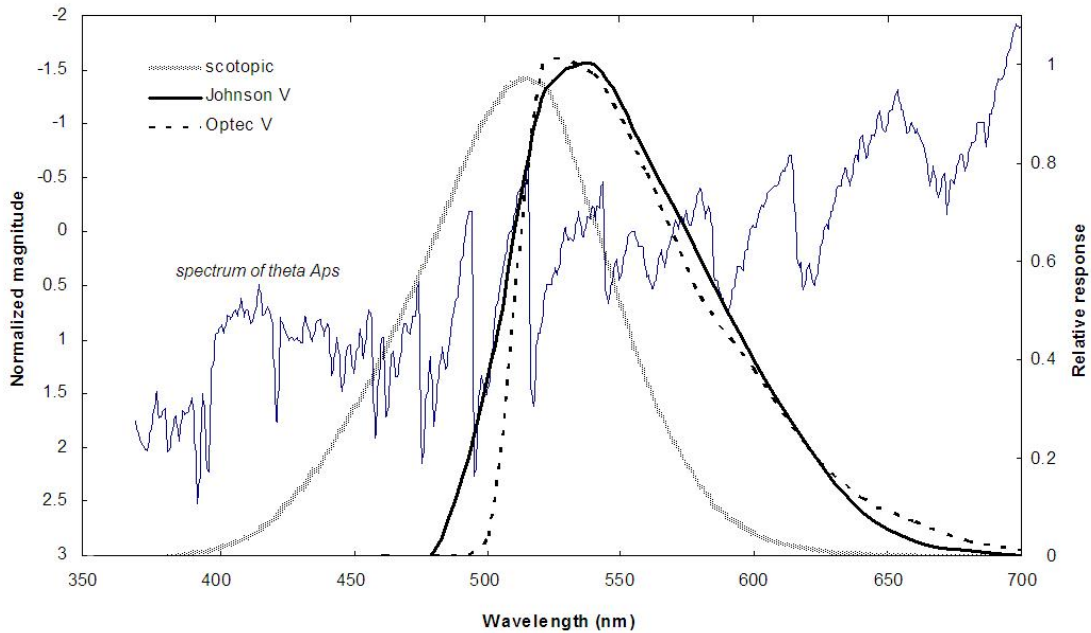

*Figure 1. Comparison of spectral responses for scotopic vision, V-band of Optec photometer and standard Johnson V-band. The spectrum of θ Aps is also shown.*

The problem is particularly complicated for SR variables owing to the prominent molecular absorption bands present in their spectra. Inspection of Figure 1 shows that some of these will be included within the spectral response for the human eye and excluded from that for the photoelectric photometer and vice versa. A small shift in wavelength of the spectral response of a detector could then make a noticeable difference in the measured magnitude.

To explore the effects of detector spectral responses and stellar spectral features on magnitudes estimated or measured, spectral responses for dark-adapted (scotopic) vision, the *V*-band of the standard *UBV* system and that for the Optec *V*-band were multiplied by the spectrum of the SR variable θ Aps (Kiehling 1987). The integrated magnitudes were then calculated. The scotopic spectral response was chosen as it represents human vision using the rod receptors in the retina of the eye – a common approach to the estimation of variable stars is to use averted vision to exploit the low-light-level sensitivity of the rods (represented by the scotopic spectral response). While this is a useful technique for estimating magnitudes for variables that change in brightness by many magnitudes and those that may be at the limit of visual detection when at their minimum brightness, for bright stars the spectral response of the human eye may be better approximated by mesopic vision – a combination of scotopic and photopic vision.

The calculated *V* magnitude using the response of the Optec photometer differed from that for the standard system by 0.02 magnitude while the magnitude calculated for scotopic vision differed by as much as 0.5 magnitude! Using cone receptors in the retina to a greater or lesser degree, this difference may be reduced as the resultant visual spectral response will be shifted to longer wavelengths and thus closer to that of the standard photoelectric *V*-band. While the techniques used by the visual observer will determine the effective spectral response for the visual observations, the calculations undertaken, here illustrate the significant effect that differences in the spectral response can have on the *V* magnitudes determined for red variables.

While experienced visual observers account for the 'Purkinje effect' (arising from the shift in the spectral response of human vision to the blue end of the spectrum at low illumination levels), small residual differences between the magnitudes of SR variables estimated visually and those measured using a photoelectric photometer are to be expected. Such discrepancies between visual and PEP observations may also be color-dependent.

**4. Transformation of PEP Measurements**
THE *UBV* standardized photometric system was introduced by Harold Johnson and William Morgan in the 1950s. Standard spectral responses for this photometric system were defined along with a set of standard stars. All measurements made with other instantiations of the *UBV* system thus require measurement of standard stars to determine the transformation relations. Provided detectors and filters are chosen carefully so that their combined responses closely match the standard system, linear transformations suffice.

While straightforward in principle, several practical problems arise:

- The standard stars do not encompass all spectral types excluding, by necessity, variable stars.
- The primary standard stars defined by Johnson and Morgan are for Northern Hemisphere observers.
- The original photomultiplier tube and filters used by Johnson and Morgan have been replaced by different brands hence later systems approximate rather than replicate the spectral responses of the original system.

These practical problems pose significant challenges for PEP measurements of bright SR variables at southern declinations. Firstly, virtually all M-type giants vary to some extent with amplitudes tending to increase for later types (Percy and Harrett 2004). Secondly, measurements of SR variables usually require extrapolation of the linear transformations determined using earlier-type stars that have smaller values of *B-V*. And thirdly, finding sufficiently 'red' comparison stars that are both bright and close by to a SR of interest can be difficult.

To minimize these problems the comparison stars chosen are K-type giants, preferably those of later type with *B-V* ~ 1.4. Fortunately these are also common enough among bright stars so that there is usually one sufficiently close to the SR variable being measured. Many bright stars have been extensively measured over the years and their magnitudes and colors well determined. The General Catalogue of Photometric Data (GCPD) is a heterogeneous source of photometric data for bright stars where multiple measurements of a star have been combined. The catalogue thus provides a useful source of well-determined magnitudes and colors for calculating transformation relations.

The question arises as to how well the *V* and *B-V* data in this catalogue, a compendium of measurements by different observers using different equipment, represents a consistent and standard system and, particularly, how well the system can be applied to the measurement of SR variables. To check this, measured *B-V* indices of 30 SR variables were compared to their GCPD values. The measurements represent a homogeneous set of data where linear relations for transformation of *V* and *B-V* have been well-established for stars ranging in *B-V* from 0 to 1.5. As is the case for all observers, this relation was then extrapolated for redder stars.

Figure 2 shows the plot of the measured *B-V* indices as a function of the GCPD values for 30 southern, bright SR variables. Also plotted is the line for a one-to-one correspondence between the *B-V* measurements and listed GCPD values. Some of the SR variables, particularly the redder ones, vary substantially in *B-V* hence error bars have been drawn showing the range of their measured *B-V* variations. GCPD values may, however, represent measurements at one part of the cycle of variation in *B-V*. Within the error bars shown the agreement is good providing confidence in the:

- fidelity with which a readily-available commercial photometer can measure *V* and *B-V* for SR variables

- consistency of *V* and *B-V* data listed in the GCPD for SR variables

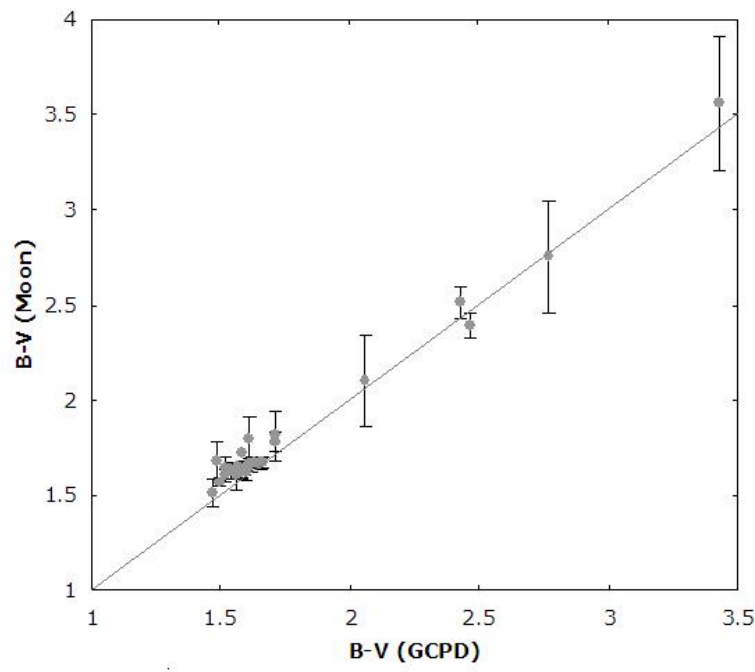

*Figure 2. Measured B-V indices as a function of the GCPD values for 30 southern, bright SR variables.*

This would also suggest that, where the photometer's spectral response is sufficiently well-matched to the standard system, the linear transformations determined using earlier-type stars may be extrapolated to redder stars such as SR variables; the resultant accuracy being determined by the differences arising from inclusion or exclusion of spectral features in the pass-band of the photometer. For an SR variable with a *B-V* ~ 1.7 (such as θ Aps), this could amount to several hundredths of a magnitude.

## 5. Visual and PEP Observations for θ Aps

Visual observations of θ Aps were made in Argentina from JD 2451621 to JD 2454180 using the Argelander method. Independently, photoelectric photometry was undertaken in Australia from JD 2452676 to JD 2454157. Subsequently, correspondence between the authors led to the pooling of their observations for further analysis.

The visual observations were made with the naked eye using a modified version of the Argelander method where the visual magnitude for a bright variable star was estimated relative to several comparison stars but using direct (cone) vision rather than averted (rod) vision as the spectral response for cone vision better approximates the response of the *V*-band. As all M-type stars are believed to be variable to some extent choosing comparison stars of similar color (*B-V* index) to SR variables is problematic. The approach taken was to try to choose comparison stars with *B-V* indices as similar (i.e. red) as possible, i.e. late-K giants. As it is not always possible to find comparison stars of similar brightness and color, and sufficiently close to the SR variable being observed, the visual observing technique developed also attempts to correct for color differences by using a different observing approach depending on the stars' color and brightness and the sky background. Typically, bright red stars will saturate the cones causing overestimation of the star's magnitude; conversely, faint red stars will not activate the cones so their brightness may be underestimated. To reduce these effects, quick glances with slightly averted vision is used in the former case and direct vision in the latter. A bright background can cause underestimation of the brightness of blue stars and overestimation of the brightness of red ones so a mix of cone and rod vision is used to minimize such effects. Using these techniques, along with photoelectric values for the magnitudes of the comparison stars rather than the values given in visual charts, estimates of the *V* magnitude of a variable are made rather than the traditional $m_{vis}$ values based on rod (scotopic) vision. A full description of the observing technique used can be found in the video and presentation slides of a talk given at the 94th meeting of the AAVSO two years ago (URL: http://www.aavso.org/aavso/meetings/fall05video/adv_otero.mov). This observing method can achieve a precision of 0.05 magnitudes (Otero, Fraser and Lloyd 2001). As indicated in Section 3, molecular absorption bands in the spectra of red giants will, however, affect the visual estimates.

The photoelectric measurements were made with an Optec SSP-5A photometer attached to a permanently-mounted 10 cm telescope housed in an observatory with a roll-off roof. For each star, 5 consecutive measurements, each of 10 seconds integration time, were taken through the *V* filter. On some nights *B*-band measurements were also taken. As the observatory is situated in an outer suburb of a major city (Adelaide), the background sky was measured for each star. When measuring through both *B* and *V* filters the sequence was $V_{star}$, $B_{star}$, $B_{sky}$, $V_{sky}$ with the time recorded at the switching of the filters during the sequence of measurements of the star through the two filters.

Measurements of θ Aps were always bracketed by measurements of two comparison stars and were usually part of more extensive observing sessions where a group of bright, southern SR variables and their nearby comparison stars were measured. This allowed atmospheric extinction to be evaluated on each night from the group of comparison stars measured. Calibrations for transformation to standard *V* magnitude and *B-V* color index have been established and are checked periodically. All comparison star values in this paper are taken from the *General Catalogue of Photometric Data* (GCPD, Mermilliod *et al.* 1997), rounded to the nearest 0.01 magnitude.

Corrections were applied to all photoelectric measurements for differences in air mass. The corrected magnitudes were then transformed to standard *V* magnitudes and *B-V* color indices. This transformation involves a correction, as a function of *B-V*, to the measurements through the *V* filter. The standard deviation for all the *B-V* measurements of θ Aps was 0.04 magnitude, however there was a large discrepancy between measured values and those listed in catalogues. The GCPD lists a *B-V* of 1.48 for θ Aps, bluer than would be expected for an M6.5III star. Consequently, the average measured value of 1.68 was used for transforming *V* filter measurements to standard *V* magnitudes. Considering the correction coefficient was 0.07, any nightly variations resulting from using the average rather than measured value (*B-V* was not measured on all nights) would typically be no more than 0.003 magnitudes.

HR 5547 was used as the primary comparison star for most of the PEP measurements with a variety of other stars used to check its constancy. (For some of the earlier measurements α Aps was used as the primary comparison star but HR 5547 was subsequently chosen as it can be measured on the same photometer sensitivity setting.) The deviations of measurements of this comparison star from its GCPD value were also monitored. Over the course of the observations presented here, the average *V* magnitude of HR 5547 was in agreement with its GCPD value to within a few thousandths of a magnitude. Standard deviations of each measurement are also calculated; the mean standard deviation of the photoelectric measurements for θ Aps being 0.01 magnitudes.

Figure 3 shows the observations made where a shift of -0.05 magnitudes has been applied to the visual estimates to bring their mean value into accord with that for the PEP measurements.

**6. Results**

6.1. Analysis by Terry Moon (using PerSea software)

As both the visual and photometric sets of data cover many cycles with a substantial interval in common, it was possible to compare them for systematic differences during the period analysis process. Analysis undertaken using the software package 'PerSea', which is based on the optimal period search method of A. Schwarzenberg-Czerny (Maciejewski 2005), gave a mean *V* from visual estimates (comprising 431 points) of 5.42 while the PEP measurements (comprising 200 points) gave a *V* = 5.37. Considering the differences in the spectral responses of the eye and the photometer as discussed in Section 4, this difference is small and may be corrected for by adding -0.05 magnitude to the visual estimates of *V*.

The interval chosen for a period search was 7 to 1000 days with the visual and photoelectric data first analyzed separately to determine the mean magnitude for each (confirming the difference of 0.05 magnitudes between the visual estimates and PEP measurements) and to gauge the difference in the period determined using only the visual or PEP observations. They were then analyzed as a single, combined dataset. A main peak found in the periodogram corresponded to a period of about 111 days. Table 2 summarizes the results from the analysis using PerSea giving the period of the dominant peak, an estimate of the precision in its determination; the mean $V$ magnitude calculated the range in the light curve and the number of points used in the analysis.

Table 2. Results from analysis using PerSea software.

|  | Period (days) | $V_{mean}$ | Range in $V$ | points |
|---|---|---|---|---|
| **Visual** | 110.6 ± 0.2 | 5.42 | 1.50 | 431 |
| **PEP** | 111.1 ± 0.2 | 5.37 | 1.52 | 200 |
| **Combined** | 111.1 ± 0.1 | 5.37 | 1.58 | 631 |

6.2. Analysis by Laszlo Kiss (using Period04 software)

The visual and PEP data was also analyzed using Period04 (2007) which is a standard approach for period searches in pulsating stars and the latest version of the original code written by Michael Breger back in the 1970s. This software allows pre-whitening in the time domain, so that, after finding the frequency of a best-fit sine-wave, that frequency is subtracted from the data and the residuals re-analyzed until there is a significant peak in their power spectrum. Period04 also offers different weighting schemes. When applied, each point was weighted by the inverse square of its stated error. As for the analysis using PerSea, periods in the order of the interval over which the observations spanned, i.e. ~ 2600 days, were considered artifacts.

First, analysis was undertaken using both weighted and non-weighted PEP data only. For both weighted and non-weighted data the three periods found were very similar, the main effects of weighting being to increase the amplitudes and reduce the scatter in the resulting fits to the data (see Table 3).

Table 3. Analysis with Period04 using PEP data only.

|  | Period (days) | | Amplitude | |
|---|---|---|---|---|
|  | weighted | *non-weighted* | weighted | *non-weighted* |
| **1 (main)** | 111.1 | *111.2* | 0.39 | *0.35* |
| **2** | 1260 | *1203* | 0.24 | *0.2* |
| **3** | 103 | *100* | 0.20 | *0.16* |

For the visual data, the stated error was mostly 0.05 and occasionally 0.1 magnitude. The results of the analyses for both weighted and non-weighted visual data are given in Table 4.

Table 4. Analysis with Period04 using visual data only.

|  | **Period (days)** | |
|---|---|---|
|  | weighted | *non-weighted* |
| **1 (main)** | 110.6 | *110.6* |
| **2** | 965 | *926* |
| **3** | 99.3 | *99.4* |

Finally, combining the PEP and visual data give results as presented in Table 5.

*Table 5. Analysis with Period04 combining the PEP and visual data.*

|  | **Period (days)** | |
|---|---|---|
|  | weighted | *non-weighted* |
| **1 (main)** | 111.0 | *110.6* |
| **2** | 1297 | *1002* |
| **3** | 101.1 | *99.3* |

6.3. Analysis by Sebastian Otero (using AVE software)

Visual and PEP data were also analyzed with the AVE software using two algorithms – PDM and Scargle (Barberá 1996). Similar to the analysis using PerSea, zero-point corrections were made to bring the mean *V* magnitude of the visual estimates into agreement with that for PEP measurements. The two datasets were then analyzed separately and combined over their interval in common, i.e. from JD 2452676 to 2454156. A predominant period of 111.2 ± 0.1 day was found in the two separate and combined analyses and using both algorithms - PDM and Scargle. Using all the available data collected since JD 2451621, both visual and PEP, a predominant period of 110.6 day was determined from both the PDM and Scargle algorithms.

**7. Discussion and Conclusions**

Figure 3 illustrates that visual and PEP data of SR variables can be successfully combined for subsequent analysis provided suitable methods are followed with making both the visual and PEP observations. Corrections ~0.05 magnitude may, however, need to be applied owing to the significant differences in the spectral responses of the human eye and photoelectric *V*-band. (See also Otero and Moon, 2006 where a similar correction is applied to the visual observations of β Gru.)

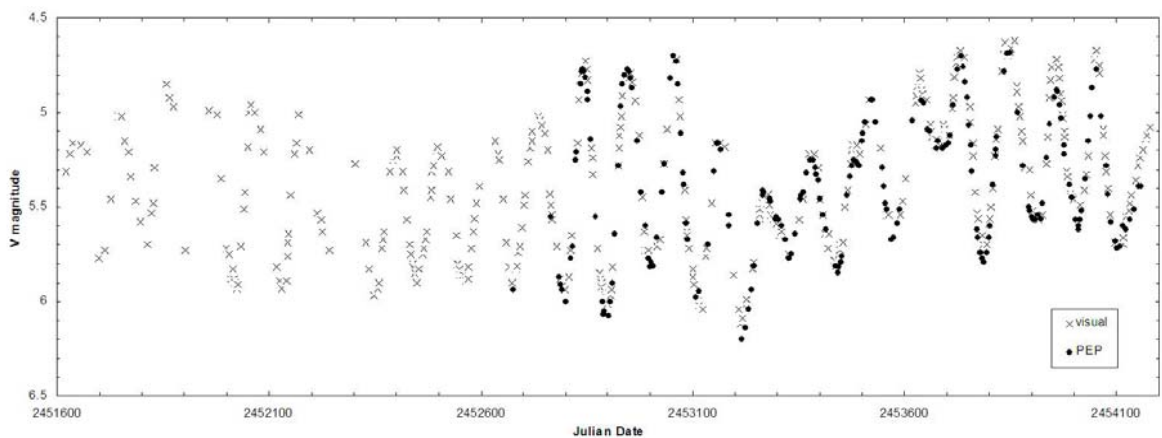

*Figure 3. PEP and Visual data obtained for θ Aps with the Visual data shifted by -0.05 magnitude.*

The advantage of combining visual and PEP observations is that it offers a potentially efficient means to monitor longer-term variations of SR variables where datasets of 100 cycles are probably needed to adequately ascertain the nature and calculate the period of any longer-term variations. Visual observations can be made by more observers, more frequently and thus can be used extend and 'fill out' the dataset obtained through more accurate photoelectric measurement. The PEP data provides, however, the means to adjust visual observations to the standard *UBV* system.

Three significantly different software packages were used to search for periodicity in the data. The similarity between the 3 sets of results obtained for θ Aps suggests that there may be no clear case for choosing one particular software package over another – the choice being mainly a matter of personal preference and familiarity with the software. Also, weighting the data did not appear to make a substantial difference to the results obtained for the predominant period and only a small difference for the possible secondary periods.

For θ Aps a predominant period of about 111 day was determined with, possibly, a longer period variation, ~ 1000 day or so, and maybe a smaller, shorter-period variation of around 100 days. The longer-period variation for θ Aps is about 10 times the predominant period; this longer-term variation of about an order-of-magnitude slower than the predominant pulsation period is observed in about 25% of semi-regular variables (Olivier & Wood 2003). The phenomenon, also known as Long Secondary Periods (LSPs), is yet to be fully explained (Wood et al. 2004); if confirmed, θ Aps is one of the brightest southern LSP variables and hence a favorable target for further detailed investigations (e.g. using interferometry).The hint of a shorter-period variation of around 100 days for θ Aps would give a period ratio of 1.1. This ratio, giving rise to 'beating' in the light curve, is also observed in other SR variables (Kiss et al. 1999). A combination of radial and non-radial oscillations may explain this phenomenon.